# Large-area printing of ferroelectric surface and super-domains for efficient solar water splitting

*Yu Tian[1,2#], Yaqing Wei[3#], Minghui Pei[1#], Rongrong Cao[4], Zhenao Gu[5], Jing Wang[6], Kunhui Liu[3], Dashan Shang[4], Jiebin Niu[4], Xiaoqiang An[7], Run Long[3*] and Jinxing Zhang[1*]*

[1]Department of Physics, Beijing Normal University, 100875 Beijing, China.

[2]School of Basic Medical Science, Air Force Medical University, 710032 Shaanxi Xi'an, China.

[3]College of Chemistry, Key Laboratory of Theoretical & Computational Photochemistry of Ministry of Education, Beijing Normal University, 100875 Beijing, China.

[4]Key Laboratory of Microelectronics Devices and Integrated Technology, Institute of Microelectronics, Chinese Academy of Sciences, 100029 Beijing, China.

[5]Key Laboratory of Drinking Water Science and Technology, Research Center for Eco-Environmental Sciences, Chinese Academy of Sciences, Beijing 100085, China

[6]Advanced Research Institute of Multidisciplinary Science, Beijing Institute of Technology, Beijing 100081, China.

[7]Center for Water and Ecology, State Key Joint Laboratory of Environment Simulation and Pollution Control, School of Environment, Tsinghua University, Beijing 100084, China.

#These authors contributed equally to this work.

*Corresponding authors email: runlong@bnu.edu.cn; jxzhang@bnu.edu.cn

**Abstract**

Surface electronic structures of the photoelectrodes determine the activity and efficiency of the photoelectrochemical water splitting, but the controls of their surface structures and interfacial chemical reactions remain challenging. Here, we use ferroelectric $BiFeO_3$ as a model system to demonstrate an efficient and controllable water splitting reaction by large-area constructing the hydroxyls-bonded surface. The up-shift of band edge positions at this surface enables and enhances the interfacial holes and electrons transfer through the hydroxyl-active-sites, leading to simultaneously enhanced oxygen and hydrogen evolutions. Furthermore, printing of ferroelectric super-domains with microscale checkboard up/down electric fields separates the distribution of reduction/oxidation catalytic sites, enhancing the charge separation and giving rise to an order of magnitude increase of the photocurrent. This large-area printable ferroelectric surface and super-domains offer an alternative platform for controllable and high-efficient photocatalysis.

**Introduction**

Photoelectrochemical (PEC) water splitting, which converts sustainable solar energy into green hydrogen fuel, provides a promising solution for the energy crisis and environmental pollution [1, 2]. Considering the three steps of the water splitting reaction (light harvesting, charge separation and catalytic reactions), the separation of the photo-generated carriers and the interfacial kinetic reactions play the critical roles in determining the efficiency of solar water splitting [3, 4]. Chemical dopants [5], heterojunction designs [6] and cocatalysts for hydrogen evolution reaction (HER) or oxygen evolution reaction (OER) [7-10] have been well developed, where the maximum efficiency is limited by the intrinsic band structures of the conventional catalyzing materials [11]. Surface modification of metal oxides with noble metals [8, 12] and defects [13] (e.g. oxygen vacancies [14, 15]) suffers from the cost of noble metals and/or chemical instability [16, 17]. Therefore, it is desirable to explore an alternative strategy for designing high-efficient, stable and controllable photoelectrodes in PEC water splitting [18, 19].

Ferroelectric semiconductors with switchable polarization provide an effective platform to control the surface/interface structures [20-22] and the kinetic processes of charge transfer [23]. The spontaneous polarization in ferroelectrics significantly enhances [23-25] and directs [26-28] the drift/separation of photo-generated electrons and holes. Benefit from the switchable polarization, the energy landscapes of ferroelectric surface and ferroelectric/electrolyte interface can be manipulated via rearranging the free charge carriers nearby the interface [29-34]. Bismuth ferrite ($BiFeO_3$; BFO) with a large polarization [32, 35] and a direct bandgap within solar visible-light spectrum ($E_g$ = 2.6−2.8 eV) [31, 36], is considered as an excellent candidate for an efficient photoelectrode. The energy bands of BFO straddle the water redox levels, due to the high-lying valence band composed of the Bi 6s and O 2p atomic orbitals, suggest its potential for efficient solar water splitting [32, 36]. Furthermore, large-scale printing of polarization and surface structure by ferroelectric/liquid ionic interaction [28] may provide an effective pathway for high-performance and controllable HER and OER on a single BFO photoelectrode, which is absent in conventional semiconductors [37].

**Polarization-dependent $H_2$/$O_2$ evolution.** Epitaxial BFO thin films were grown on (001) $SrTiO_3$ (STO) substrates with conductive $(La,Sr)MnO_3$ (LSMO) layer for PEC measurement. In this work, the thickness of the BFO thin films is 50 nm, which leads to a pronounced photocurrent density due to the carrier diffusion length [32]. The detailed growth, piezoresponse force microscopy (PFM) and PEC measurements can be seen in Supporting Information. Fig. 1 (a, b) shows that BFO thin films with upward (downward) polarization give a cathode (anode) photocurrent. This polarization-selective photocathode and photoanode are benefited from its special energy bands (straddle the water redox levels) as illustrated in Fig. 1 (c, d). The upward-polarized BFO induces positive (negative) bond charges at the electrolyte/BFO (BFO/LSMO) interface, and downward (upward) band bending towards the electrolyte/BFO (BFO/LSMO) interface [32]. This band bending promotes the photo-generated electrons to BFO surface, and thus favors the HER (Fig. 1 (c)). While, the downward polarization gives rise to the upward (downward) band bending towards the electrolyte/BFO (BFO/LSMO) interface, which promotes the holes to BFO surface, and facilitates the OER (Fig. 1 (d)). These controllable polarization and ferroelectric surface suggest that photocathode and photoanode may be assembled on a single photoelectrode.

**Hydroxyl-bonded BFO surface enhanced HER and OER simultaneously.** Reconstruction of surface chemical structure of the photoelectrode may increase the energy bands offset and thus facilitate charge transfer across the photoelectrode/electrolyte interface [38]. Thus, we design the BFO surface structure through a controllable ferroelectric/water ionic interaction (Supporting Figure 1) [28]. A dramatic enhancement of the photoemission peak (~531.7 eV) in the X-ray photoelectron spectroscopy (XPS) spectra after the ionic interaction indicates the presence of terminal hydroxyls (metal-oxide-hydrogen, M-O-H) on the BFO surface [39, 40] as shown in Fig. 2 (a). These hydroxyls gradually desorbed from the surface when the sample was annealed at elevated temperature (Supporting Figure 2), which is similar with previous observations [41, 42], indicating that the emergent surface structure (M-O-H) is hydroxyl-bonded ferroelectric surface (BFO-OH). In water splitting process, surface hydroxyl groups may play a significant role in interfacial charge transfer [43, 44]. Interestingly, the BFO-OH photoelectrodes with high photo-stability (Supporting Figure 3) not only drive hydrogen evolution with a photocurrent density of -0.06 mA·cm$^{-2}$ at 0 V vs. RHE (two times larger than pristine BFO), but also catalyze oxygen production with a photocurrent density of 0.07 mA·cm$^{-2}$ at 1.23 V vs. RHE as shown in Fig. 2 (b), which does not exist in the pristine BFO with the same polarization (Fig. 1 (b)).

A low charge transfer resistance in the Nyquist plot of the BFO-OH (Fig. 2 (c)) implies a new process for electrons/holes transfer across the interface, which may benefit from a reconstructed surface electronic structure in the BFO-OH (no morphology changes before and after the hydroxyl modification, Supporting Figure 4).

The surface electronic structure of the BFO-OH was characterized by scanning tunneling spectroscopy (STS) [45], showing a positive shift of conduction band minimum (CBM) and valance band maximum (VBM) as seen in Fig. 2 (d). Density functional theory (DFT) calculations indicate that the surface density of states of BFO-OH, where the OH$^-$ ions bond to the Fe sites on the FeO$_2$-ternimated energy-favorable BFO (001) surface [28, 46, 47], is consistent with the STS results, as shown in Fig. 2 (e). The up-shift of both the CBM and VBM of the BFO-OH in Fig. 2 (f) ascribes to the formation of a built-in field from hydroxylated surface to the bottom of the photoelectrode [28], leading to the decreased barriers for electrons and holes migration at the same surface, as illustrated in Supporting Figure 5. The details of DFT calculation and STS measurement are presented in Supporting Information.

**Dynamic process of water splitting on the BFO-OH.** To understand the dynamic reaction processes of the enhanced HER and OER on the BFO-OH, we calculated the Gibbs free energy of each step in the HER and OER and plotted the energy diagram in Fig. 3 (a) and 3 (b), respectively. In general, HER follows a two-electron step:

$* + H^+ + e^- \rightarrow *H$ (Eq.1)

$*H + H^+ + e^- \rightarrow H_2$ (Eq.2)

where the symbol "*" represents the active sites of catalysts. Due to a rational thermokinetics balance between the above two reaction steps, the Gibbs free energy for *H intermediate generation ($|\Delta G_{H*}|$) on the BFO-OH decreases to 0.39 eV, decreasing by a factor of 3 compared to the pristine BFO (1.46 eV), which notably enhances the HER efficiency, as illustrated in Fig. 3 (a). In contrast, OER is a four-electron process:

$* + H_2O \rightarrow *OH + H^+ + e^-$ (Eq.3)

$*OH \rightarrow *O + H^+ + e^-$ (Eq.4)

$*O + H_2O \rightarrow *OOH + H^+ + e^-$ (Eq.5)

$*OOH \rightarrow O_2 + H^+ + e^-$ (Eq.6)

Production of oxygenated intermediates *OH in step 1 (depicted by Eq. 3) on the BFO-OH only need to overcome low energy barrier of 1.6 eV, as shown in Fig. 3(b), resulting from adsorbed water molecules dissociation followed by the Fe-OH-OH formation, which favors driving the subsequent reactions, and leading to an efficient activity for OER [38, 48, 49]. However, the Gibbs free energy of production of *OH in step 1 on the pristine BFO, where the hydroxyls bound to the surface Fe atoms, is as high as 2.18 eV (Supporting Figure 6), which makes the subsequent processes rarely occur and the OER stops. The overpotential η in OER decreases to 0.53 V in the BFO-OH from 0.86 V in the pristine BFO (Supporting Information), suggesting an improved OER efficiency in the BFO-OH due to the terminal hydroxyls induced facilitated charge transfer and decreased water adsorption energy. During the OER, hydroxyls in the BFO-OH are acting as the catalytic sites other than the intermediates during the OER, which is supported by secondary ion mass spectroscopy (SIMS) in Fig. 3 (c, d). BFO-OD was designed by deuterium oxide treatment, as validated by the depth profiles (Supporting Figure 7). The intensity of deuterium on the BFO-OD almost remains identical before and after the PEC measurements in aqueous solution (Fig. 3 (d)), indicating the robust bonds (BFO-OD/OH) act as stable active sites for capturing the photo-generated electrons and holes for driving water splitting.

**Printing of ferroelectric super-domains for efficient water splitting.** To further enhance the water splitting efficiency, a large-area printing of checkboard down/up domains with up/down depolarization fields ($E_{dp}$) on the BFO-OH are schematically demonstrated in Fig. 4 (a), where a typical 500 nm periodic super-domains structure is shown in Fig. 4 (b). A decrease of the domain size leads to an enhanced and stable photocurrent density of OER and HER, as shown in Supporting Figure 8. It is notable that the photocurrent density dramatically increases by one order of magnitude when the domain size reduces from 1 mm to 500 nm as shown in Fig. 4 (c). DFT calculations were performed for understanding the mechanism of the enhanced efficiency on the checkboard domains, where the spatial approach of the charge densities of the VBM and CBM (supported by oxygen and iron respectively) on the uniform upward-polarized BFO with downward depolarization field favors the enhancement of the electrons-phonons coupling and accelerates charge recombination. The antiparallel polarization on the BFO-OH with antiparallel depolarization field separates the electrons/holes effectively with reduced charge recombination by localizing the CBM and VBM at different domains as shown in Fig. 4 (d), and thus leads to a high current density. We expect a higher current density if the domain size reduces to the limit of electrons/holes diffusion length of the bifunctional ferroelectric photoelectrode, which may further figure out the mismatched pH for photocathode and photoanode in integrated systems (Supporting Figure 9) [37], and builds up a platform for controllable and high-efficient photocatalysis with photo- oxidation and reduction coupling reaction [50, 51].

**Conclusions**

In this work, a ferroelectric photoelectrode with an efficient electrons and holes migration during PEC reaction is fabricated by large-area construction of hydroxyls bonding on BFO surface. This robust hydroxyls bonding on the BFO lifts the surface CBM and VBM, decreasing the formation energy of oxygenated/hydrogenated intermediates, and leading to a simultaneously enhanced efficiency of HER and OER for water splitting in this bifunctional photoelectrode. Furthermore, the charge separated surface band edge states on the BFO-OH through large-area printing of ferroelectric super-domains with spontaneous up/down checkboard electric fields, further reduce the recombination of the photo-generated electrons and holes, providing a new paradigm for designing controllable surface (interface) structures and steering charge transfer for high-efficient catalysis.

**Competing interests:** The authors declare no competing interests.

**Acknowledgements**: JZ acknowledges the support from the National Key Research and Development Program of China (2016YFA0302300) and CAS Interdisciplinary Innovation Team. JZ also acknowledges the sup-port from the National Natural Science Foundation of China (11974052), Beijing Natural Science Foundation (Z190008). R. L. acknowledge support of the National Foundation of China, grant no. 51861135101, the Recruitment Program of Global Youth Experts of China and the Beijing Normal University Startup.

**Figures and captions**

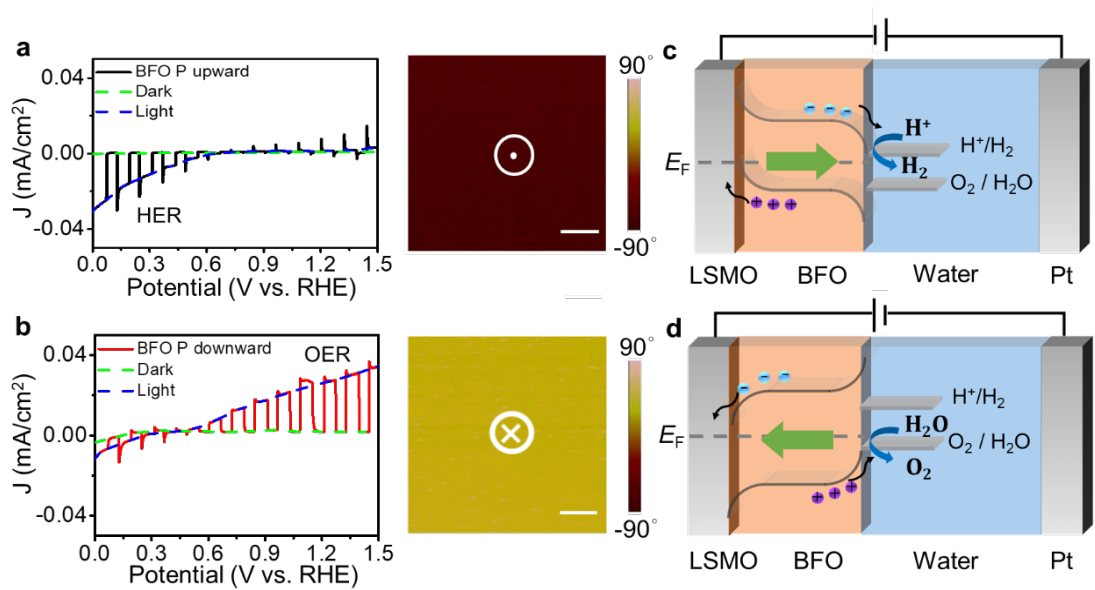

**Fig. 1 Polarization-dependent HER and OER on ferroelectric photoelectrode.** PEC response and PFM image for upward- (**a**) and downward-polarized (**b**) BFO thin films respectively. Schematics of band diagram during PEC water splitting for the case of upward- (**c**) and downward-polarized (**d**) BFO thin film respectively. Scale bar 2 μm in (**a**, **b**).

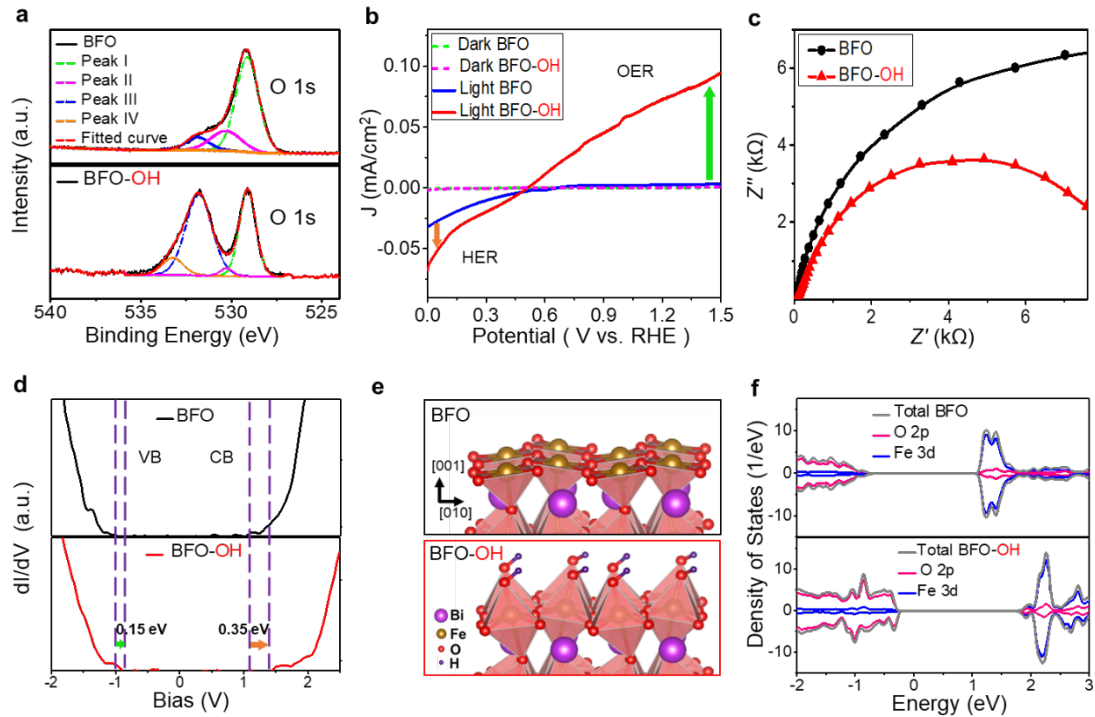

**Fig. 2 Surface design (BFO-OH) for simultaneously enhanced HER and OER. (a)** XPS spectra for the surface of BFO and BFO-OH respectively. The O 1s XPS spectra for the BFO and BFO-OH surface can be fitted into four components: peak I (~529.1 eV, oxygen in the perovskites), peak II (~530.3eV, oxygen vacancies or lowly charged oxygen particles ($O^-$)), and peak III (~531.7eV, hydroxyl groups on BFO), peak IV (~533.2eV, absorbed water molecules) [40, 41, 47]. (**b**) Current-voltage curves for BFO and BFO-OH thin films. (**c**) Nyquist plots for BFO and BFO-OH recorded at 0 V vs. RHE. STS (**d**), structure diagrams (**e**) and calculated surface density of states (**f**) for BFO and BFO-OH.

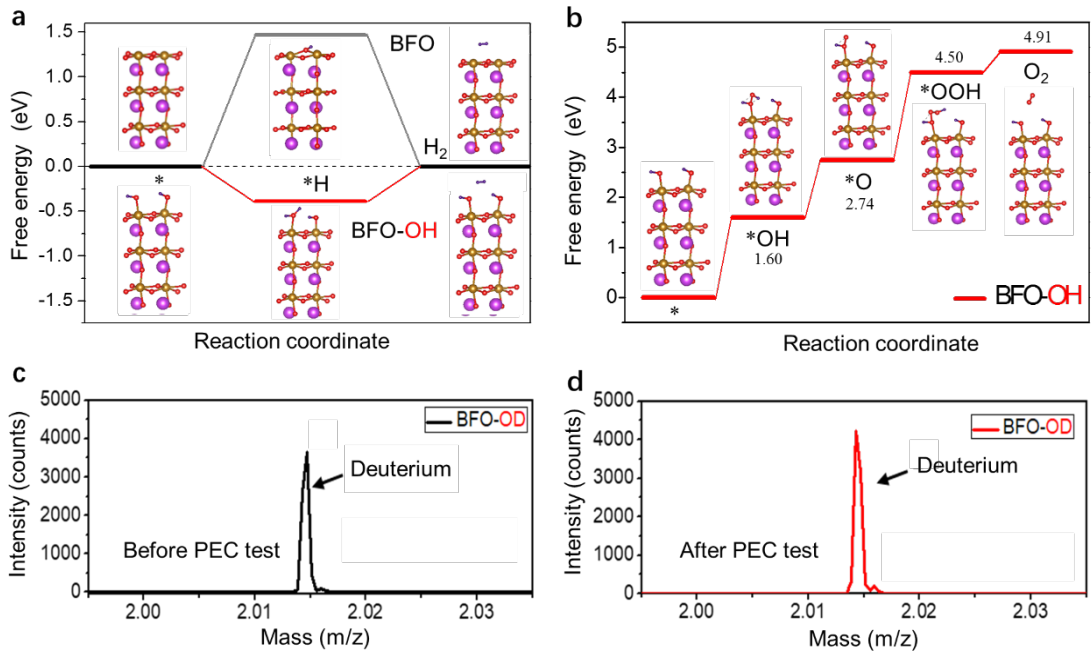

**Fig. 3 Dynamic process of the efficient water splitting on the BFO-OH.** (**a**) Free energy diagram of HER for BFO and BFO-OH. (**b**) Free energy diagram of OER for BFO-OH, where the OER starts with the dissociation of the adsorbed water molecule into *OH (energy barrier of 1.6 eV), and then dehydrogenating into *O (1.14 eV), reacting with another water molecule forming *OOH (1.76 eV) and dissociating to oxygen (0.41 eV). SIMS of deuterium on the BFO-OD film before (**c**) and after (**d**) the PEC measurements.

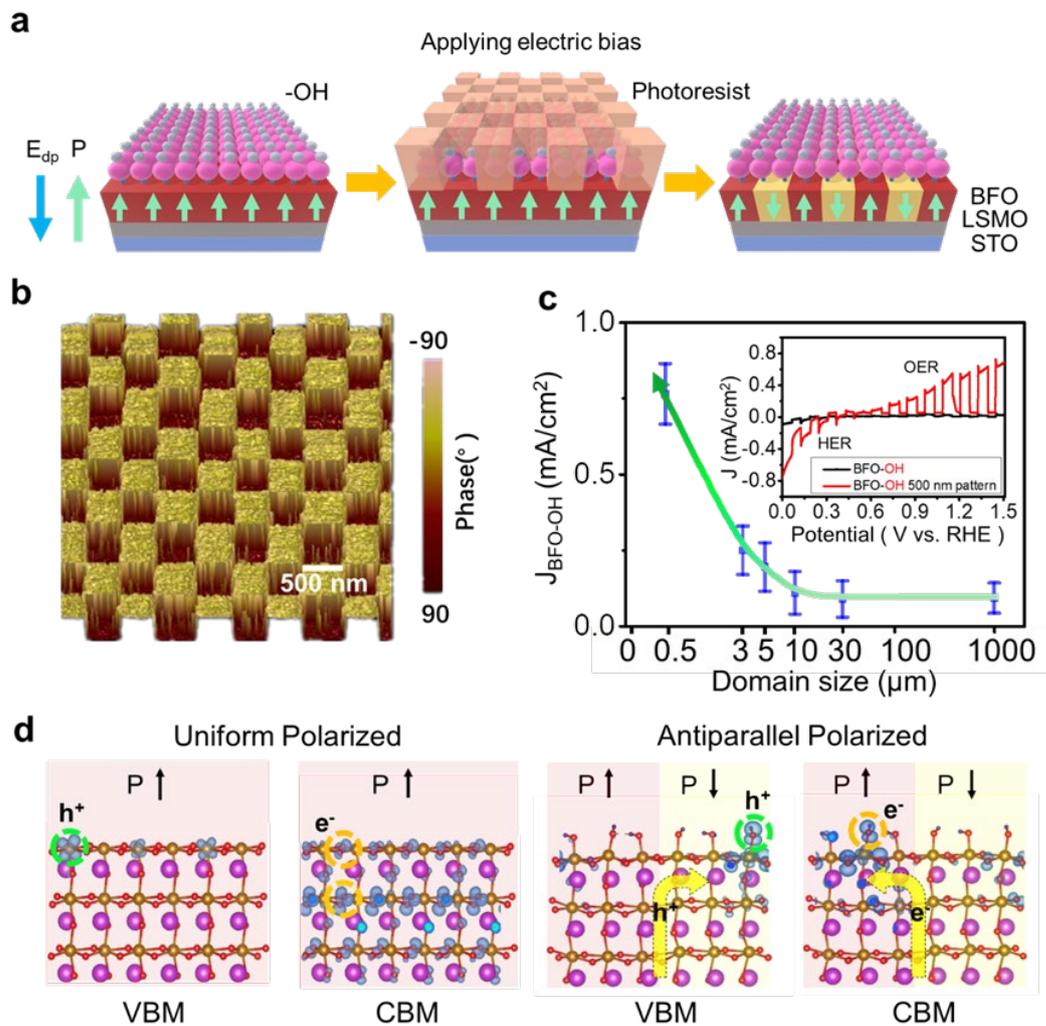

**Fig. 4 Printing of ferroelectric super-domains for efficient water splitting.** (**a**) Printing process of ferroelectric super-domains with checkboard up/down depolarization fields. The sequence involves modifying the BFO surface with hydroxyls via water treatment, then creating a patterned coating on the BFO surface by photoengraving, switching the polarization of exposed region by applying external electric field and removing the photoresist. The green (blue) arrows describe the out-of-plane component of the polarization ($E_{dp}$). (**b**) PFM image of BFO-OH thin film with 500 nm periodic super-domains. (**c**) The current density of the BFO-OH with super-domains significantly increases with the decrease of the domain size, inset shows the current density of the BFO-OH and BFO-OH with 500 nm periodic super-domains. (**d**) Charge densities of VBM and CBM for the pristine BFO and BFO-OH with antiparallel polarization.